\pdfoutput=1
\documentclass[twocolumn,PRB,showpacs,amsmath,amssymb,superscriptaddress,floatfix]{revtex4}

\usepackage{bm}
\usepackage{graphicx}
\usepackage{amssymb}
\usepackage{amsmath}
\usepackage{xcolor}

\begin{document}

\title{Evolution of Non-Kramers Doublets in Magnetic Field in PrNi$_2$Cd$_{20}$ and PrPd$_2$Cd$_{20}$}

\author{A. M.  Konic}  
\affiliation{Department of Physics, Kent State University, Kent, Ohio, 44242, USA}

\author{R. B. Adhikari}
\affiliation{Department of Physics, Kent State University, Kent, Ohio, 44242, USA}

\author{D. L.  Kunwar}  
\affiliation{Department of Physics, Kent State University, Kent, Ohio, 44242, USA}

\author{A. A. Kirmani}
\affiliation{Department of Physics, Kent State University, Kent, Ohio, 44242, USA}

\author{A. Breindel}
\affiliation{Department of Physics, University of California at San Diego, La Jolla, CA 92903, USA}
\affiliation{Center for Advanced Nanoscience, University of California, San Diego, La Jolla, California 92093, USA}

\author{R. Sheng}
\affiliation{Department of Physics, University of California at San Diego, La Jolla, CA 92903, USA}
\affiliation{Center for Advanced Nanoscience, University of California, San Diego, La Jolla, California 92093, USA}

\author{M. B. Maple}
\affiliation{Department of Physics, University of California at San Diego, La Jolla, CA 92903, USA}
\affiliation{Center for Advanced Nanoscience, University of California, San Diego, La Jolla, California 92093, USA}

\author{M. Dzero}
\affiliation{Department of Physics, Kent State University, Kent, Ohio, 44242, USA}

\author{C. C. Almasan}
\affiliation{Department of Physics, Kent State University, Kent, Ohio, 44242, USA}

\date{\today}

\begin{abstract}
Praseodymium-based 1-2-20 cage compounds Pr$T_2X_{20}$ ($T$ is generally Ti, V, Nb, Ru, Rh, Ir; and $X$ is either Al, Zn or Cd) provide yet another platform to study non-trivial electronic states of matter ranging from topological and magnetic orders to unconventional multipolar orders and superconductivity. In this paper, we report measurements of the electronic heat capacity in two Pr-based 1-2-20 materials: PrNi$_2$Cd$_{20}$ and PrPd$_2$Cd$_{20}$. We find that the lowest energy multiplet of the Pr $4f^2$ valence configuration is a $\Gamma_3$ non-Kramers doublet and the first excited triplet is assumed to be a magnetic $\Gamma_5$. By analyzing the dependence of the energy splitting between the ground and first excited singlet states on external magnetic field, we found that the maximum in the heat capacity corresponding to the Schottky anomaly in PrPd$_2$Cd$_{20}$, unlike PrNi$_2$Cd$_{20}$, shows pronounced linear dependence on external magnetic field at higher field values. This effect is associated with the exchange interactions between the field-induced magnetic dipole moments.
\end{abstract}

\maketitle
\section{Introduction}
The discovery of the second-order mean-field-like phase transition in URu$_2$Si$_2$ provides a remarkable example of an ordered phase with an unknown order parameter, hence the name 'hidden-order transition'. Since its discovery almost thirty five years ago \cite{HOMydosh1985,Maple1986,Schlabitz1986,Review2014}, quite significant experimental and theoretical progress has been made to get insights into the microscopic mechanism governing this transition (see Ref. \onlinecite{Mydosh2020} for the most recent review). In particular, two singlet states of the ground state valence configuration $5f^2$ of the uranium ions seem to be a key feature to take into account in trying to identify the symmetry of the hidden order state \cite{KH2009,KH2010}. 

Naturally, one may wonder whether the 'hidden order' state could emerge in other materials with partially filled $f$-orbitals \cite{Patri1,MultiTheory}. Based on the body of knowledge accumulated for URu$_2$Si$_2$, the pre-requisite for the 'hidden-order' state seems to be that valence configuration of the $f$-orbital multiplet should have an even number of electrons (non-magnetic) and have a non-Kramers doublet as its ground state. At low enough temperatures, this allows one to represent the $f$-states on each ion in terms of two-level systems. Despite the fact that it is not possible to directly couple to the 'hidden-order' parameter, most likely a multipolar one, interactions between these two-level systems may then lead to  either a 'hidden-order' or antiferromagnetic state, depending on the relative strength of the corresponding exchange parameter \cite{KH2010,Patri1}.

Relatively novel 1-2-20 cage compounds with the chemical formula Pr$T_2X_{20}$ ($T$ is generally Ti, V, Nb, Ru, Rh, Ir; and $X$ is either Al, Zn or Cd) \cite{Onimaru2010,Onimaru2012,Yazici2015} appear to satisfy these criteria. These materials are remarkable for a fairly strong hybridization between the conduction and $f$-electron states \cite{Satoru1,Satoru2}. Furthermore, at very low temperatures some of these materials seem to develop some type of long-range order: PrIr$_2$Zn$_{20}$ and PrRh$_2$Zn$_{20}$ develop superconductivity at $T\sim 0.05$ and $0.6$ K, respectively \cite{Onimaru2010}, PrTi$_2$Al$_{20}$ shows a ferroquadrupole order at 2 K \cite{Satoru1,aquadro}, PrRh$_2$Zn$_{20}$ an antiferroquadrupole order at 0.06 K, while the order is still undetermined in PrV$_2$Al$_{20}$ \cite{Satoru1,Onimaru2012,Freyer1,Patri1}.

The Pr ion in Pr$T_2X_{20}$ ($X=$Zn, Al) finds itself in the $4f^2$ (Pr$^{3+}$) valence configuration with total angular momentum $J=4$. The nine-fold degeneracy is lifted by the crystalline electric field resulting in a $\Gamma_{1}$ singlet, a non-Kramers $\Gamma_{3}$ doublet, and $\Gamma_{4}$ and $\Gamma_{5}$ triplet states also seen in similar Pr-based compounds \cite{Yazici2015,Onimaru2012,Kusunose2016}. Moreover, thermodynamic measurements seem to indicate that the lowest multiplet is a non-Kramers $\Gamma_3$ doublet \cite{Onimaru2012,WHITE2015} which, as has already been noted above, opens the way for emergence of the exotic multipolar ordered phases \cite{Flint1,Flint2}. 

PrNi$_2$Cd$_{20}$ and PrPd$_2$Cd$_{20}$ are relatively new additions to the family of the 1-2-20 materials. Previous analysis on heat capacity measurements performed down to 2 K on PrNi$_2$Cd$_{20}$ and PrPd$_2$Cd$_{20}$ gave indication of a $\Gamma_{3}$ non-Kramers doublet ground state, with the energy gaps of 12 K and 11 K between the ground state and the triplet excited state for PrNi$_2$Cd$_{20}$ and PrPd$_2$Cd$_{20}$,  respectively. In addition neither sample show any signs of ordering down to temperatures of $T\sim 0.02$ K \cite{Yazici2015}. It has subsequently been shown through ultrasonic measurements that the ground state of the Pr ion in the PrNi$_{2}$Cd$_{20}$ material is a $\Gamma_{3}$ non-Kramers doublet \cite{Yanagisawa2020}. In this context, we are motivated by not only confirming this recent finding by using heat capacity measurements for both PrNi$_{2}$Cd$_{20}$ and PrPd$_{2}$Cd$_{20}$, but also, through Schottky fits of the specific heat curves, we extracted the the Sommerfeld coefficient $\gamma$ and the values of the energy-level splitting between the two singlets that make up the doublet ground state as well as between the doubly-degenerate ground state and the triply-degenerate first-exited state.  In addition, we show how the splitting between the singlets of the non-Kramers doublet varies with external magnetic field and estimate the strength of the exchange interactions in these compounds. Hence, this would predict what state - superconductivity or some kind of multipolar order - these materials would develop first upon further cooling into the millikelvin range. 

In this paper, we report specific heat measurements done on single crystalline samples of PrNi$_2$Cd$_{20}$ and PrPd$_2$Cd$_{20}$ in external magnetic fields. Fits of the specific heat data at temperatures below 4 K, and entropy calculations from those fits were done to further investigate the ground states of these systems. We found that the ground state of Pr ion is, indeed, the non-Kramers $\Gamma_{3}$ doublet. Furthermore, the predominantly linear dependence of the peak in the heat capacity on external magnetic field in PrNi$_2$Cd$_{20}$ suggests that, on one hand, the energy between the ground state doublet and first excited triplet state is small enough for the triplet state to be completely excluded from the analysis and, on the other hand, the interaction between the induced dipole moments lead to boosting the linear-in-field contribution.

\section{Experimental Details}
Single crystals of PrNi$_2$Cd$_{20}$ and PrPd$_2$Cd$_{20}$ were grown at the University of California San Diego using the Cd self-flux technique described in Refs.~\cite{Yazici2015,Burnett2014}. Analysis of the powder x-ray diffraction (XRD) patters obtained via a Bruker D8 Discover x-ray diffractometer was done to determine the crystal structure and quality of the single crystals \cite{Yazici2015}. This analysis showed both samples to be single phase crystals lacking any indication of impurity phases \cite{Yazici2015}. The structure for both samples was determined to be the CeCr$_{2}$Al$_{20}$-type cubic structure, having a space group of $Fd\Bar{3}m$ \cite{Yazici2015,Burnett2014}.

We performed heat capacity measurements on a PrNi$_{2}$Cd$_{20}$ sample with a mass of 0.9 mg and a PrPd$_{2}$Cd$_{20}$ sample with a mass of 1.1 mg, using the He-3 insert for a Quantum Design Physical Property Measurement System (PPMS) employing a standard thermal relaxation technique.  For better thermal contact between the samples and measurement platform, the contact surface of each sample was polished with sand paper. These measurements were performed in magnetic fields $B$ ranging from 0 to 14 T applied along the $c$-axis, i.e., perpendicular to the single crystals and over a temperature range of 0.39 K $\le T \le 50$ K.

\section{Results and Discussion}
\begin{figure}[ht!]
	\centering
	\includegraphics[width=1.05\linewidth,height=.85\linewidth]{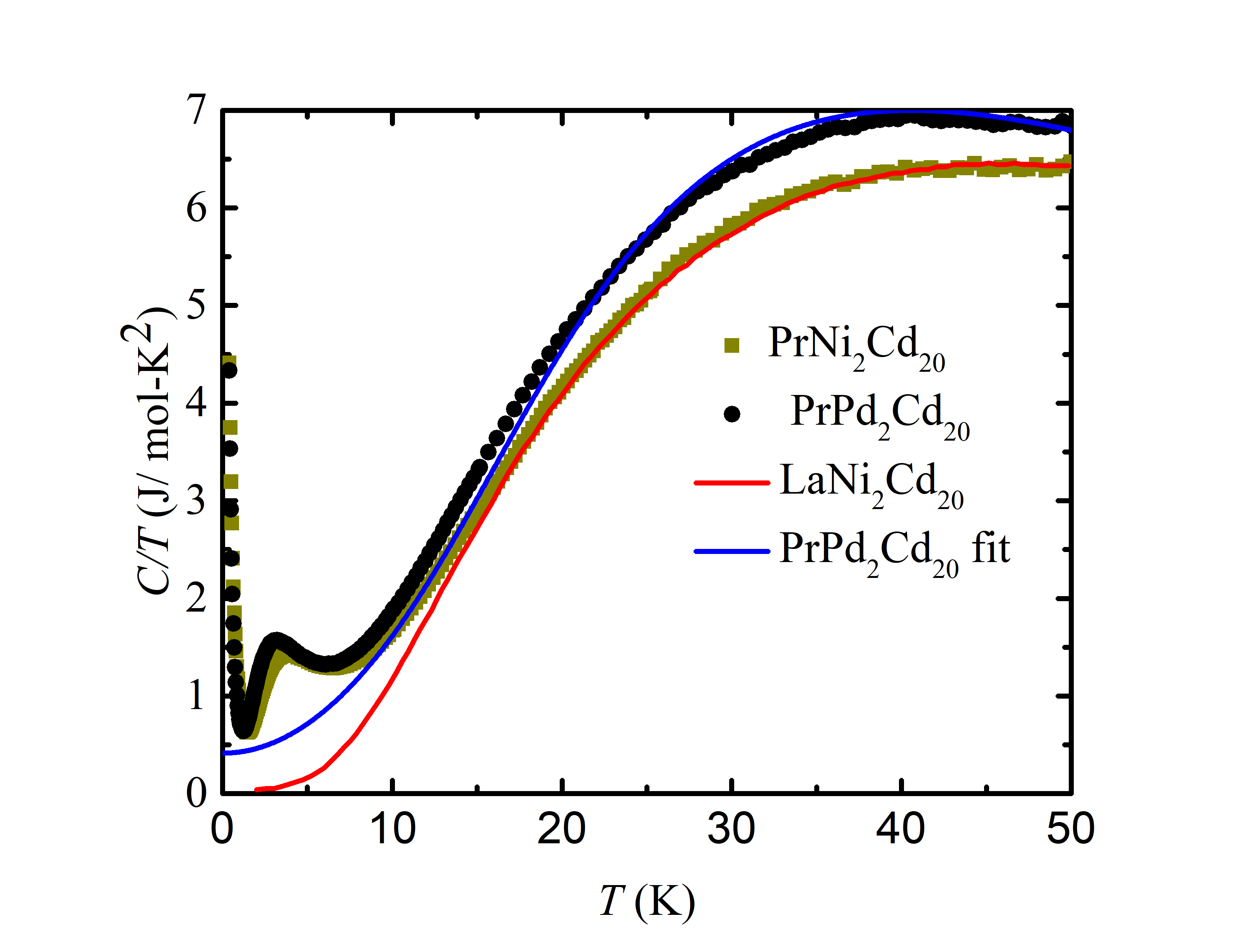}
	\caption{Temperature dependence of specific heat normalized by temperature, $C/T$, for both PrNi$_{2}$Cd$_{20}$ (green squares) and PrPd$_{2}$Cd$_{20}$ (black circles). To further analyze the $4\textit{f}$ electrons contribution, the non-magnetic compound LaNi$_{2}$Cd$_{20}$ was used to subtract the phonon contribution to the specific heat for PrNi$_{2}$Cd$_{20}$, while the phonon contribution for PrPd$_{2}$Cd$_{20}$ is determined as discussed in the text.}\label{Fig1}
\end{figure}

Figure~\ref{Fig1} depicts the temperature dependence of the measured specific heat normalized by temperature, $C/T$, for single crystalline samples of PrNi$_2$Cd$_{20}$ and PrPd$_2$Cd$_{20}$ obtained in zero magnetic field. As can be seen, both compounds exhibit a clear 
Schottky-type peak just below 5 K, with additional upturns seen below 1 K. To further investigate these features of the specific heat, we examined the $4f$ electrons contribution to the specific heat in more detail. 

We extract the $4f$ electrons contribution to the specific heat of PrNi$_2$Cd$_{20}$ by subtracting the specific heat of the non-magnetic analog compound LaNi$_2$Cd$_{20}$ (data also shown in Fig.~\ref{Fig1}, red curve) from the specific heat of PrNi$_2$Cd$_{20}$ over the whole temperature range, as has been previously done \cite{Morrison1968}. Due to the overlap between the data for PrNi$_2$Cd$_{20}$ and LaNi$_2$Cd$_{20}$ in the higher temperature range, the LaNi$_2$Cd$_{20}$ data can be subtracted from the PrNi$_2$Cd$_{20}$ data directly.

The phonon contribution to the heat capacity for the PrPd$_2$Cd$_{20}$ compound had to be obtained by fitting the data over the temperature range $20 \leq T \leq 50$ K with the standard expression
\begin{equation}
C(T)=\gamma T+{9{ N}k_B}t^3\int\limits_0^{\frac{1}{t}}\frac{x^4e^{x}dx}{(e^x-1)^2},
\end{equation}
where $\gamma$ is the Sommerfeld coefficient, ${N}$ is the number of primitive cells per mole, ($9{N}k_B\approx 1720$ $\textrm {J}/{(\textrm{mole}\cdot\textrm{K}})$), $k_B$ is Boltzmann constant, and $t=T/\theta_D$ ($\theta_D$ is the Debye temperature, which for PrPd$_2$Cd$_{20}$ equals approximately $145$ K).
\begin{figure}[ht!]
	\centering
		\centering
		\includegraphics[width=1\linewidth]{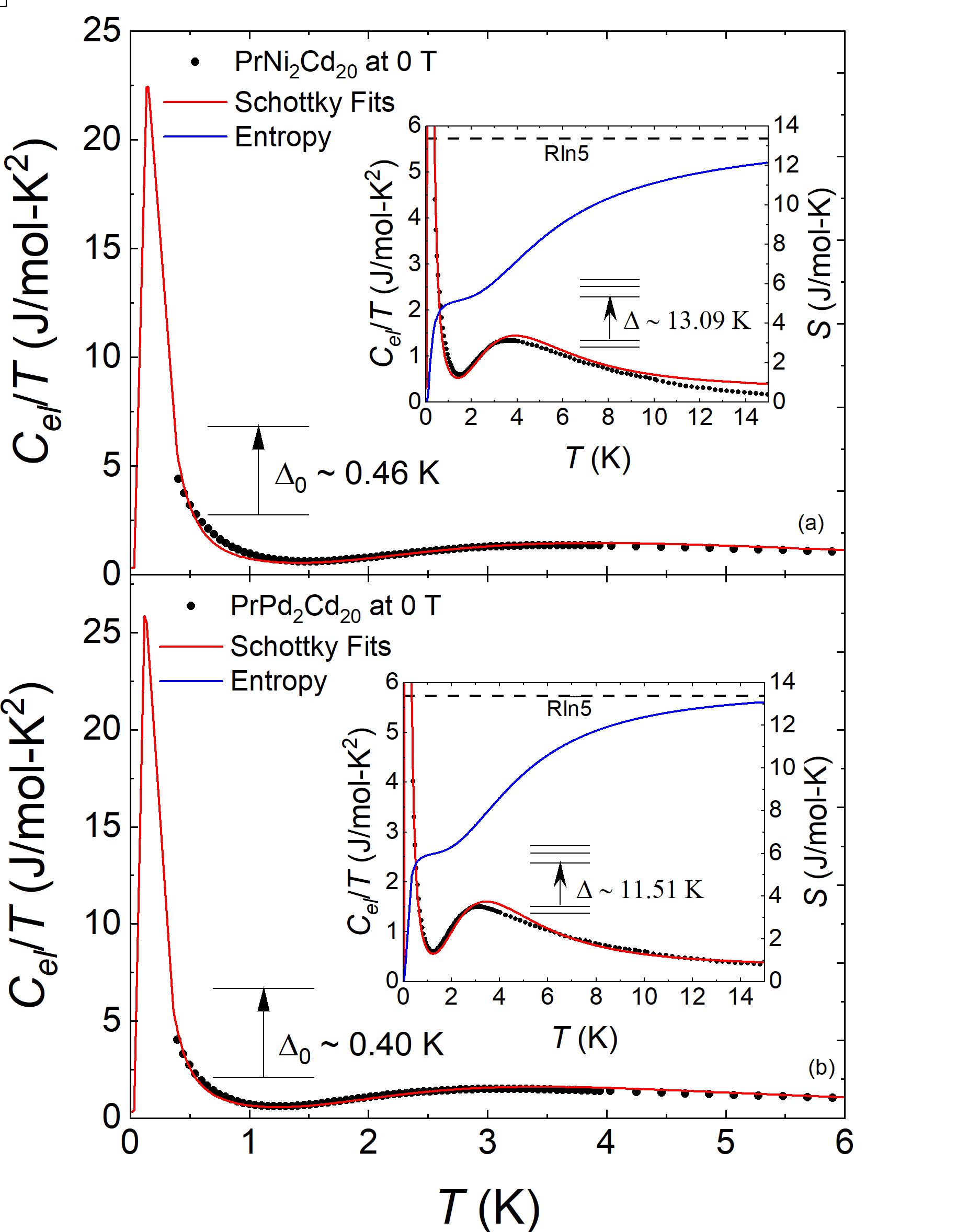}
		\caption{Electronic specific heat normalized by temperature, $C_{el}/T$, (left axis) and entropy $S$ (right axis insets) as a function of temperature $T$ for (a) PrNi$_{2}$Cd$_{20}$ and (b) PrPd$_{2}$Cd$_{20}$. The main panels focus on the low $T$ upturns, while the insets reveal the Schottky-type peaks at higher $T$. Fits of the Schottky-type peaks (red lines) in both the main panels and insets were done using a two-level model (see text for details).}\label{Fig2}
\end{figure}
\begin{figure}[ht!]
		\includegraphics[width= 1\linewidth]{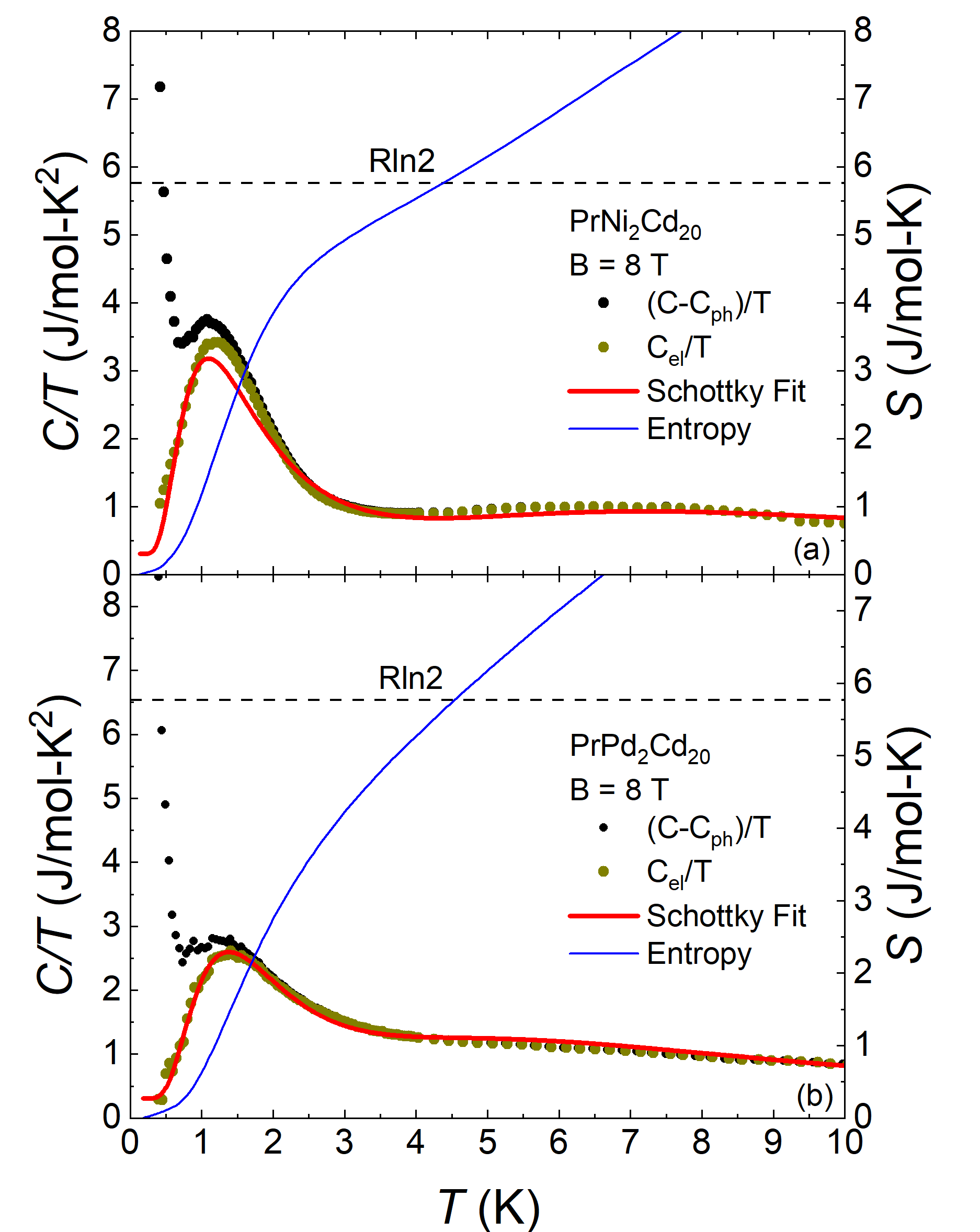}
		\caption{Specific heat normalized by temperature, $(C-C_{\textrm{ph}})/T$ and $C_{el}/T$, (left axis) and entropy $S$ (right axis) vs temperature $T$ data measured in an applied magnetic field $H$ of 8 T for (a) PrNi$_{2}$Cd$_{20}$ and (b) PrPd$_{2}$Cd$_{20}$. $C_{el}$ is the electronic specific heat obtained by subtracting the nuclear $C_n$ contribution from $C-C_{\textrm{ph}}$, with $C_n \propto T^{-2}$. The Schottky peak seen just above 1 K is a result of the shift in the peak location from $T< 0.3$ K, in response to the 8 T field. Note that the entropy exceeds the $R\ln 2$ value signalling the field-induced admixture of the first excited triplet. }
		\label{Fig3}
		\end{figure}
Then, to obtain the electronic contribution to the specific heat, we have also subtracted the nuclear $C_n$ contribution to the specific heat from $C-C_{\textrm{ph}}$, with $C_n \propto T^{-2}$ for $T\lesssim1$ K and magnetic fields $\gtrsim 4$ T, where the nuclear contribution becomes significant. 

Next, we determined the energy level splitting and the entropy to further study the ground state structure of the $4f$ multiplet, as presented below. Figures~\ref{Fig2}(a) and ~\ref{Fig2}(b) and their insets depict, for both compounds, the temperature dependencies of  $C_{\textrm{el}}/T$ on the left axis and of the entropy $S$ on the right axis. We fitted the peaks near 4 K (data and fits shown in the insets of the two figures), and the upturns below 1 K (data and fits shown in the main panel of the figures) using a two-level Schottky model given by \cite{Yazici2015}:
\begin{equation}
		\frac{C_{\textrm{el}}}{T} = \gamma +
		\frac{A\Delta^2g_ag_be^{-\Delta/T}}{T^3\left(g_a+g_be^{-\Delta/T}\right)^2}.
\end{equation}
Here $A\leq 1$ is a phenomenological parameter which reflects the degree of the hybridization between Pr ions and the conduction band, $\Delta$ is the energy level splitting, and $g_a$ and $g_b$ are the degeneracy of the ground and first excited state, respectively. In what follows, we use $\gamma$, $A$, and $\Delta$ as  fitting parameters.

By fitting the Schottky peaks present in the two insets of Figs.~\ref{Fig2}(a) and ~\ref{Fig2}(b), we obtain values for the energy splitting between the ground state and first excited state. We obtained the best fits with $g_a=2$ and $g_b=3$. This implies that the ground state is a non-Kramers doublet and the first excited state is a triplet. We obtained $\gamma = 0.30 \pm 0.01$ and $\Delta\ = 13.09 \pm 0.16$ K for PrNi$_2$Cd$_{20}$, and $\gamma = 0.41 \pm 0.01$ J/mol-K$^{2}$ and $\Delta\ = 11.51 \pm 0.09$ K, for PrPd$_2$Cd$_{20}$, with $A = 1$ for both compounds. These values are in good agreement with values reported previously \cite{Yazici2015}. The fact that in both of these systems $A\ = 1$ implies sufficiently weak hybridization between the conduction and f electron states. This is consistent with the non-Kramers doublet nature of the ground state multiplet. We note that the values of $A$ and $\gamma$ for both compounds in zero magnetic field were held constant for the subsequent fitting analysis done in higher magnetic fields. 

\begin{figure}[ht!]
	\centering
		\centering
		\includegraphics[width= .99\linewidth]{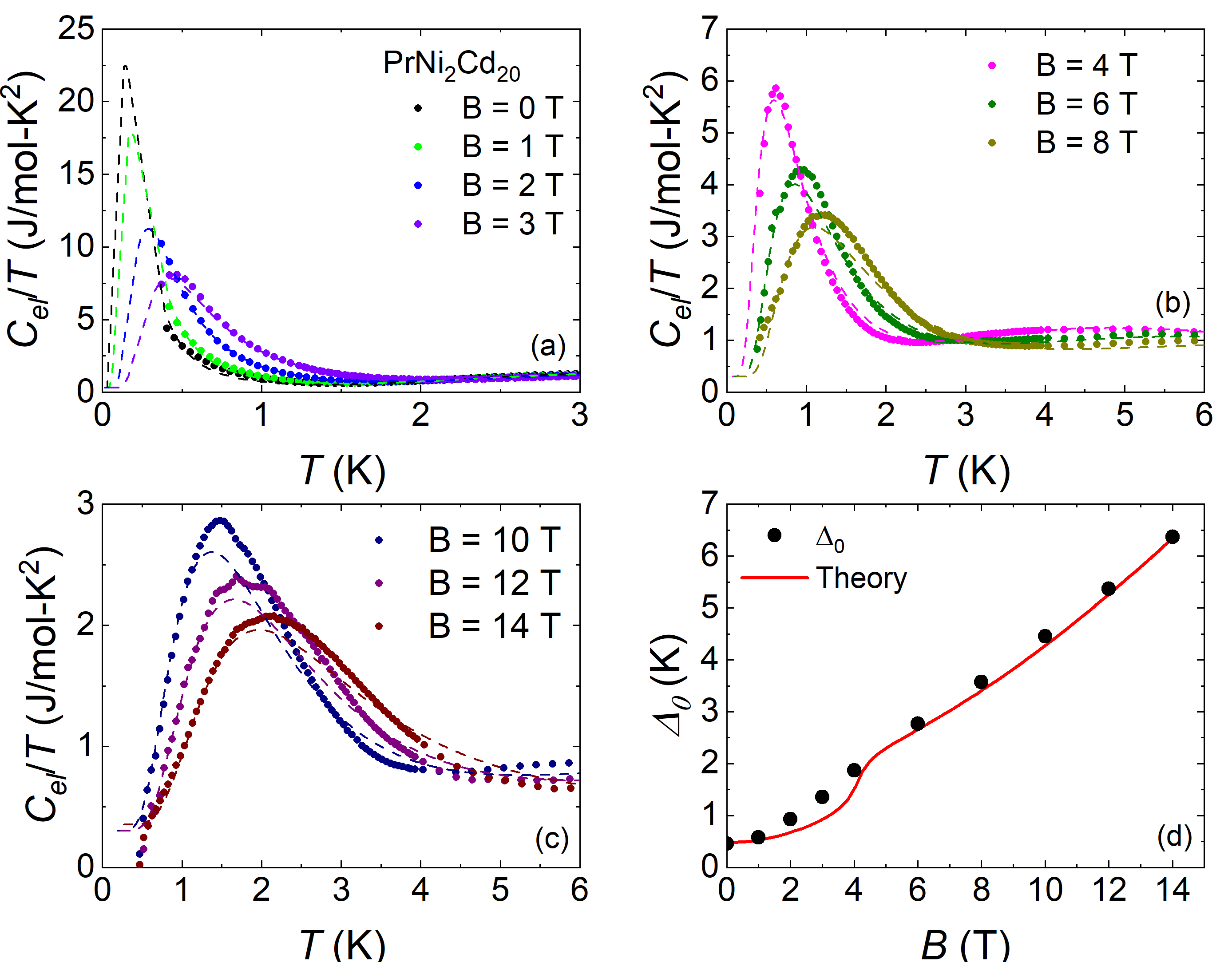}
		\caption{(a-c) Electronic specific heat normalized by temperature $C_{el}/T$ vs temperature $T$ and Schottky fits (dashed lines) in applied magnetic fields $H$ for PrNi$_{2}$Cd$_{20}$  . (d) Energy gap $(\Delta_{0})$ between the two ground state levels, as a function of magnetic field $B$.}
		\label{Fig4}
\end{figure}

Next, we assume that the Schottky upturns in specific heat seen below 1 K (main panel of Figs.~\ref{Fig2}(a-b)) for each compound is a result of the lifting of the degeneracy of the non-Kramers doublet ground state. Hence, we fitted these data with Eq. (2) with  $g_{a} = g_{b} = 1$ and obtained values for the energy splitting between the two levels of the ground state to be, $\Delta_{0} = 0.461 \pm 0.006$ K and $\Delta_{0} = 0.401 \pm 0.004$ K for PrNi$_{2}$Cd$_{20}$ and PrPd$_{2}$Cd$_{20}$, respectively. We note that we obtained worse fits of these data by choosing other values of $g_{a}$ and $g_{b}$, hence other ground state options. These results, therefore, further confirm that the ground state of both compounds is the non-Kramers doublet.

In Figures~\ref{Fig3}(a) and ~\ref{Fig3}(b) we show the specific heat for both compounds measured in an applied magnetic field of 8 T. These plots exhibit how $C_{el}/T$ data (dark yellow points) were obtained by subtracting a low $T$ nuclear contribution from $(C-C_{ph})/T$ (black data points). Subsequently, we fitted these latter data measured in different magnetic fields. These results reveal that the zero-field upturns seen in Figs.~\ref{Fig2}(a) and ~\ref{Fig2}(b) below 1 K shift to higher temperatures with increasing applied magnetic field and the peaks that are not visible in $B=0$ down to 0.3 K become visible at, e.g., $T \approx 1$ K in a magnetic field $B=8$ T. Fits of these Schottky peaks with Eq. (2) with  $g_{a} = g_{b} = 1$ are shown on the figures in red. At this magnetic field we obtained $\Delta_0 = 3.57 \pm 0.04$ K for PrNi$_{2}$Cd$_{20}$ and $\Delta_0 = 4.48 \pm 0.02$ for PrPd$_{2}$Cd$_{20}$. 

In order to further check the validity of the conclusions we have drawn from our fits in Figs.~\ref{Fig2}(a) and ~\ref{Fig2}(b) and their insets, we also calculated the entropy from the data at a magnetic field of $B = 8$ T where the peak that develops is clearly visible, using 
\begin{figure}[ht!]
	\centering
		\centering
		\includegraphics[width= .99\linewidth]{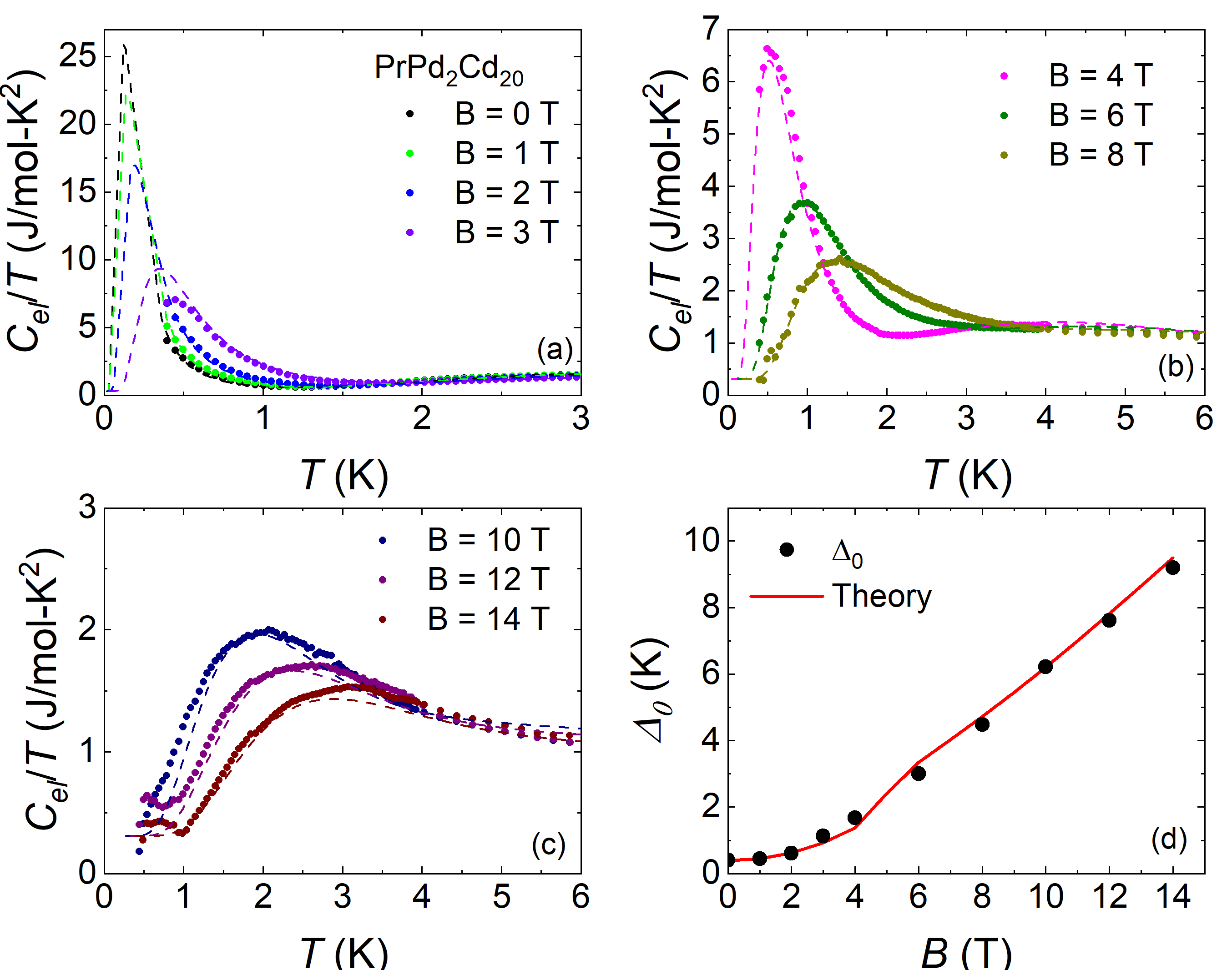}
		\caption{(a-c) Electronic specific heat normalized by temperature $C_{el}/T$ vs temperature $T$ and Schottky fits (dashed lines) in applied magnetic fields $H$ for PrPd$_{2}$Cd$_{20}$  . (d) Energy gap $(\Delta_{0})$ between the two ground state levels, as a function of magnetic field $B$.}
		\label{Fig5}
\end{figure}

\begin{equation}\label{Entropy}
S(T) = \int\limits_0^T [C_{\textrm{el}}(T)
-\gamma T]\frac{dT}{T},
\end{equation}

see blue lines in Figs.~\ref{Fig3}(a-b). The entropy was calculated using the electronic specific heat data (dark yellow data points in Fig.~\ref{Fig3}) with the extrapolation below 0.39 K made through the Schottky fit as stated before (red lines in Fig.~\ref{Fig3}). Here, both compounds achieve the expected $R\ln2 = 5.76$ J/mol$\cdot$K at T $\approx$ 5 K. This confirms that, indeed, the Schottky peak observed in 8 T for both compounds is a result of the splitting of the ground-state $\Gamma_{3}$ non-Kramers doublet. An entropy larger than $R\ln2$ at $T > 5$ K is a result of the excited triplet states contributing at the higher temperatures.

In addition, we note that the higher energy multiplets, specifically the triplet state above the ground state doublet, becomes accessible as the temperature increases. As with the lower temperature Schottky peak, we expect the overall change in entropy after this multiplet is accessed to be equal to $R\ln(g_a+g_{b})$, where now $g_{a} = 2$, and $g_{b} = 3$. Thus, we expect a change in entropy equal to $R\ln 5=13.4 $ J/mol K. As can be seen in the insets of Figs.~\ref{Fig2}(a) and \ref{Fig2}(b), we obtained $90.5\%$ of the expected value, i.e., $S= 12.13$ J/mol K for PrNi$_{2}$Cd$_{20}$ at $T = 15$ K and $97.7\%$, $S= 13.09$ J/mol K for PrPd$_{2}$Cd$_{20}$ at $T = 15$ K. Thus, we find that in this temperature range the contribution to the entropy comes from the non-Kramers doublet along with the two lowest lying states in the first excited triplet. 

We plot the temperature dependence of $C_{\textrm{el}}/T$ data  for PrNi$_2$Cd$_{20}$ and PrPd$_2$Cd$_{20}$ measured in applied magnetic fields from 0 to 14 T in Figs.~\ref{Fig4} and ~\ref{Fig5}, respectively. We note that the data for $B =$ 10, 12, and 14 T for PrNi$_2$Cd$_{20}$ (Fig. 4(c)) deviate from the Schottky fits between about 1 and 3 K and that a sharp peak appears at around 1.7 K in $B$ = 12 T. This latter feature may indicate a meta-magnetic transition, when the lowest lying state of the first excited triplet moves below the non-Kramers doublet. It also could be the reason for the fact that the data at these higher field values deviate from the Scottky fits. Nevertheless, the presence of this extra feature in the specific heat data clearly does not affect our main findings in the present work. 

We also note that, even though the ground state valence configuration of the Pr ions is non-magnetic, the location of the peaks in Figs. 4 and 5 exhibit a dependence on magnetic field. This dependence for both materials can be understood using a simple model of interacting two-level systems in an external magnetic field, which we discuss in the theoretical model section below.

By fitting the specific heat data of Figs.~\ref{Fig4} and ~\ref{Fig5} with Eq. (2) where $g_a=g_b=1$, we extract the values of the energy splitting $\Delta_{0}$ between the two levels of the non-Kramers doublet ground state as a function of magnetic field for the two compounds. We show these results in Figs. ~\ref{Fig4}(d) and ~\ref{Fig5}(d). The red solid lines are fits of these data using a phenomenological model, as discussed in the next section. 

\section{Phenomenological Model and Discussion}
In order to fit the experimental data for the dependence of the energy splitting on external magnetic field seen in Figs.~\ref{Fig4}(d) and ~\ref{Fig5}(d), we consider the ground state non-Kramers doublet with state vectors
\begin{equation}\label{Gamma3}
\begin{split}
&\vert\Gamma_3,a\rangle=\sqrt{\frac{7}{24}}\left(\vert 4\rangle+\vert-4\rangle\right)-\sqrt{\frac{5}{12}}\vert0\rangle, \\
&\vert\Gamma_3,b\rangle=\frac{1}{\sqrt{2}}\left(\vert 2\rangle+\vert-2\rangle\right).
\end{split}
\end{equation}
Our data indicate that there is a small energy splitting $\Delta_0$ between $\vert\Gamma_3,a\rangle$ and $\vert\Gamma_3,b\rangle$ states. 
For the first excited triplet state we use the results of Ref. \cite{Yazici2015} and assume that it is described by the $\Gamma_5$ triplet which lies at the energy $\Delta$ above $\Gamma_3$:
\begin{equation}
\begin{split}
&\vert\Gamma_5,a(b)\rangle=\sqrt{\frac{7}{8}}\vert\pm3\rangle-\sqrt{\frac{1}{8}}\vert\mp1\rangle, \\ &\vert\Gamma_5,c\rangle=\frac{1}{\sqrt{2}}\left(\vert 2\rangle-\vert-2\rangle\right).
\end{split}
\end{equation}
Given that in our experiments the magnetic field is along the $c$-axis, ${\vec B}\vert\vert[001]$, one can immediately check that it will induce virtual transitions between $\vert\Gamma_3,b\rangle$ and $\vert\Gamma_5,c\rangle$ states. Assuming that magnetic field $H<\Delta/\mu_B$, these virtual transitions lead to decrease in the energy of the $\vert\Gamma_3,a\rangle$ state by  $\delta \varepsilon\approx(\mu_BH)^2/\Delta$, while the energy of the 
$\vert\Gamma_5,c\rangle$ state increases by the same amount. Furthermore, the energy of the $\vert\Gamma_5,a\rangle$ decreases linearly with increasing magnetic field. 

With these provisions, we fit the dependence of the peak in the heat capacity on the magnetic field at low temperatures using a model with three energy levels $\varepsilon_1$, $\varepsilon_2$
and $\varepsilon_3$. The first two energy levels correspond to the non-Kramers doublet states: $\varepsilon_1=-\Delta_0/2-\rho\left[\sqrt{\Delta^2+(\mu_BH)^2}-\Delta\right]$ and $\varepsilon_2=+\Delta_0/2$. The third energy level corresponds to $\vert\Gamma_5,a\rangle$ state, $\varepsilon_3=\Delta-\alpha\mu_BH$. We consider $\alpha$ and $\rho$ as the fitting parameters. The best fits for the experimental data shown in Figs. \ref{Fig4}(d) and \ref{Fig5}(d) were obtained for $\Delta_0/2\approx 0.495$ K and $\rho\approx 2.95$. The values of the remaining parameters $\Delta$ and $\alpha$ are: $\Delta=10$ K, $\alpha=0.1$ for PrPd$_2$Cd$_{20}$ and $\Delta=12$ K, $\alpha=1.15$ for PrNi$_2$Cd$_{20}$. Note that the values for both $\Delta_0$ and $\Delta$, which we have chosen independently for the fits, are in agreement with those extracted from the heat capacity. Significantly higher values of the parameter $\alpha$ in PrNi$_2$Cd$_{20}$ are likely due to enhanced dipole-dipole interactions in this material. 

Lastly, we note that in principle the very similar results could be found under the assumption of $\Gamma_4$ being the first excited triplet. In order to verify that our assumption of $\Gamma_5$ being the first excited triplet indeed holds, more detailed studies of the dependence of the Schottky peak on various directions of the external magnetic field will have to be carried out. 

An analysis of this $H$ dependence of the energy splitting of the ground state shows that the exchange interactions between the two-level systems are weak in PrNi$_2$Cd$_{20}$, but they cannot be neglected in PrPd$_2$Cd$_{20}$. These latter results, therefore,  suggest that in the millikelvin range of temperatures PrNi$_2$Cd$_{20}$ could develop superconductivity, while PrPd$_2$Cd$_{20}$ will develop long-range order that could be either multipolar or magnetic.

\section{Conclusions}
We performed specific heat measurements of PrNi$_2$Cd$_{20}$ and PrPd$_2$Cd$_{20}$ in magnetic fields ranging from 0 to 14 T and in temperatures ranging from 50 K down to 0.39 K. Schottky fits of the specific heat curves show that in both compounds studied the ground state is  the non-Kramers $\Gamma_{3}$ doublet, while the first excited state is a triplet. Entropy calculations provide further evidence. Through Schottky fits of the specific heat curves we also extracted the values of the energy level splitting between the two singlets that make up the doublet ground state and between the doubly-degenerate ground state and the triply-degenerate first-exited state. We also obtained the magnetic field dependence of the energy level splitting of the ground state. 

\section{Acknowledgments} 
The work at Kent State University was supported by the National Science Foundation grants NSF-DMR-1904315, NSF-DMR-2002795, and by the U.S. Department of Energy, Office of Science, Office of Basic Energy Sciences under Award No. DE-SC0016481. The work at UCSD was supported by the US Department of Energy, Office of Basic Energy Sciences, Division of Materials Sciences and Engineering, under Grant No. DE-FG02-04ER45105 (single crystal growth) and the National Science Foundation under Grant No. NSF-DMR-1810310 (materials characterization). 
\newpage
\bibliography{Ref_Pr}
\end{document}